\documentclass[aip,apl,reprint]{revtex4-1}

\usepackage{graphicx}   
\usepackage{amssymb} 
\usepackage{amsmath}

\begin{document}
\title{Healing of a hole in a carbon nanotube under electron irradiation in HRTEM}

\author{Irina V. Lebedeva}
\email{liv_ira@hotmail.com}
\affiliation{CIC nanoGUNE BRTA, San Sebasti\'an 20018, Spain}
\author{Andrey M. Popov}
\email{popov-isan@mail.ru}
\affiliation{Institute for Spectroscopy of Russian Academy of Sciences, Troitsk, Moscow 108840, Russia}
\author{Andrey A. Knizhnik}
\email{andrey.knizhnik@gmail.com}
\affiliation{Kintech Lab Ltd., 3rd Khoroshevskaya Street 12, Moscow 123298, Russia}

\begin{abstract}
Healing of a hole in a carbon nanotube under electron irradiation in HRTEM at room temperature is demonstrated using molecular dynamics simulations with the CompuTEM algorithm. Formation of an amorphous patch is observed in all simulation runs. The amorphous patch is formed in the absence of external carbon adatoms only via reconstruction of the carbon bond network. It consists mainly of 5-, 6- and 7-membered rings and causes a small bottleneck. In addition, further growth of the initial amorphous patch under electron irradiation takes place. Two-coordinated atoms are found to play a crucial role in the latter process, analogous to autocatalisys of rearrangements of rings in fullerenes. The principal rearrangements in the presence of two-coordinated atoms can be described as generalized sp-defect migration: a bond is broken between two three-coordinated atoms and one of them forms a new bond with a nearby two-coordinated atom. If the new and former two-coordinated atoms are not bonded, the reaction leads both to displacement of the sp defect and changes in rings of the sp$^2$ carbon structure. Migration by hopping of two-coordinated atoms and other reactions involving simultaneous breakage of two bonds are also detected but much rarely. Long-living two-coordinated atoms in the patch structure and related fast growth of the patch are observed in more than half of the simulation runs. Since the amorphous patch and bottleneck affect the electronic properties of the nanotube, such nanotubes can be perspective for nanoelectronic applications.
\end{abstract}
 
\maketitle

\section{Introduction}
Healing of graphene and carbon nanotube (CNT) structure \cite{Song2011,Zan2012,Xu2012,Robertson2013,Chen2013,Ajayan1998,Kotakoski2011,Robertson2012,Girit2009,Krasheninnikov2002,Jiang2010,Kotakoski2007,Botari2016,Lee2014,Trevethan2014,Santana2013a,Van2017,Maheshwari2012,Pierlot2014} has been extensively investigated in recent years since the presence of structural defects can significantly influence electronic properties of graphene- and CNT-based devices (see, for example, Ref. \onlinecite{Banhart2011} for a review). High-resolution transmission electron microscopy (HRTEM) studies show two types of healing processes in carbon nanostructures. The first process is filling of vacancies \cite{Song2011,Robertson2013,Chen2013} and holes \cite{Zan2012,Xu2012} created in graphene by electron impacts with carbon adatoms,
which originate from existing carbon contaminations or atoms knocked out by electrons with formation of the perfect graphene lattice  \cite{Song2011,Zan2012,Robertson2013,Chen2013} or an amorphous graphene-like structure \cite{Zan2012,Xu2012,Robertson2013}. The second process of graphene \cite{Kotakoski2011,Robertson2012} and CNTs \cite{Ajayan1998} healing in HRTEM is saturation of dangling bonds via reconstruction of the carbon bond network resulting in formation of an amorphous structure. Such a process occurs through motion and coalescence of monovacancies created by electron impacts. Formation of the straight zigzag edge via migration of atoms along the edge in HRTEM \cite{Girit2009} should be also mentioned. Healing of CNTs by laser irradiation \cite{Van2017} and heating \cite{Maheshwari2012,Pierlot2014} has been also demonstrated by Raman spectroscopy but no direct information about atomic-scale structure is available from these studies.

As for atomistic simulations, the following healing processes taking place at high temperature have been considered: formation of the perfect structure as a result of the monovacancy and adatom annihilation for CNTs \cite{Krasheninnikov2002,Kotakoski2007} or via filling of holes by adatoms for graphene\cite{Botari2016} and dangling bond saturation via motion and coalescence of vacancies in graphene \cite{Lee2014,Trevethan2014} and CNTs \cite{Ajayan1998,Krasheninnikov2002,Jiang2010,Lee2014} leading to formation of the bottleneck \cite{Jiang2010}. Graphene healing through directional motion of single vacancies to the edge has been proposed \cite{Santana2013a}. Hole healing in CNTs through structural reconstruction in the absence of a source of carbon adatoms has not yet been studied. Here we consider this possibility for a single-wall CNT (SWCNT) under electron irradiation in HRTEM at room temperature and study the atomistic mechanism of the healing. Note that atomistic mechanisms of bond rearrangements in curved sp$^2$ carbon structures are of general interest for diverse processes of formation and transformation of fullerenes and CNTs, particularly for healing of CNTs by laser \cite{Van2017} and thermal \cite{Maheshwari2012,Pierlot2014} treatment.

The paper is organized as follows. First we descibe the methods used. Then we give the data obtained on hole healing, structural evolution after hole healing, yields of different CNT structures with time and atomistic mechanisms of structural rearrangements. Finally we discuss the results and summarize our conclusions. 

\section{Methods}
To determine the atomistic mechanism of healing and analyze the structure of the patch, we have performed molecular dynamics (MD) simulations modeling the action of electron irradiation through the CompuTEM algorithm \cite{Santana2013,Skowron2013}.
This algorithm allows to rescale the structure evolution time to the experimental conditions in HRTEM explicitly taking into account  momentum transfer from incident electrons to atoms  and annealing of the structure between irradiation-induced reactions. The excellent  agreement between the processes of structure transformation taking place experimentally in HRTEM and in MD simulations using CompuTEM algorithm was observed for formation of endohedral metallofullerenes from nickel clusters surrounded by amorphous carbon \cite{Sinitsa2017} and CNT cutting by a nickel cluster \cite{Lebedeva2014}. 

We have studied healing of a hole in a CNT under electron irradiation in 30 simulation runs which corresponded to the experimental times of 2800 -- 6600 s. The (5,5) CNT of 4.3 nm length was considered. A hole in the CNT wall was created by removal of 6 adjacent atoms in the CNT center (Figure \ref{fig:structure}a). 20 atoms at each end of the CNT were fixed. The in-house code MD-kMC \cite{Knizhnik2017} was used. The Brenner potential reparameterized to reproduce the energies of different graphene edges, barrier for vacancy migration and formation energy of carbon chains \cite{Sinitsa2018,Sinitsa2019} was applied. This potential gives the barriers for reactions at edges of a narrow graphene nanoribbon \cite{Sinitsa2018} and migration of two-coordinated atoms (sp defects) in fullerenes \cite{Sinitsa2020} in qualitative agreement with density functional theory calculations.

Following the CompuTEM algorithm, the structure was equilibrated at 300 K during 5 ps before each electron impact. Then the impact was modeled at 300 K for 5 ps more. If the impact was ``successful", i.e. produced any structural change, the structure relaxation before the next impact, which takes several seconds at room temperature, was modeled at temperature 2000 K during 30 ps. The temperature  was controled using the Berendsen thermostat \cite{Berendsen1984}.  The kinetic energy of electrons was 80 keV. This energy is slightly below the experimental value of the threshold energy for knock-on damage of a perfect isolated SWCNT of 86 keV \cite{Smith2001}. The threshold energy of atom displacement calculated for the (5,5) CNT considered here is 19 eV \cite{Krasheninnikov2005,Banhart2005}. This is greater than the maximal energy of 15.8 eV that can be transferred from electrons with the kinetic energy of 80 keV to carbon atoms, i.e. the (5,5) CNT should be stable under the 80 keV electron irradiation  \cite{Krasheninnikov2005}. Such a stability of the perfect (5,5) CNT structure was confirmed in the MD simulations performed. 
The electron flux was $4\cdot10^{6}$ $e^{-}$nm$^{-2}$s$^{-1}$ and it was directed perpendicular to the hexagon removed to create the hole. The probablities of electron impacts to different atoms were computed based on the minimal transferred energy assigned to each atom depending on its local surrounding. For two-coordinated atoms, atoms in non-hexagonal rings and neighbouring ones it was chosen at 13 eV and for one-coordinated atoms at 5 eV. These parameters were tested in our previous paper on CNT cutting  \cite{Lebedeva2014}. 
For the analysis of bond topology, it was assumed that two carbon atoms are bonded if the distance between them does not exceed 1.8 \AA. The hole was considered as healed when the number of atoms along its perimeter became less than 10 (initially this number was 18).

\section{Results}
\subsection{Hole Healing}

\begin{figure*}
   \centering
    \includegraphics[width=0.85\textwidth]{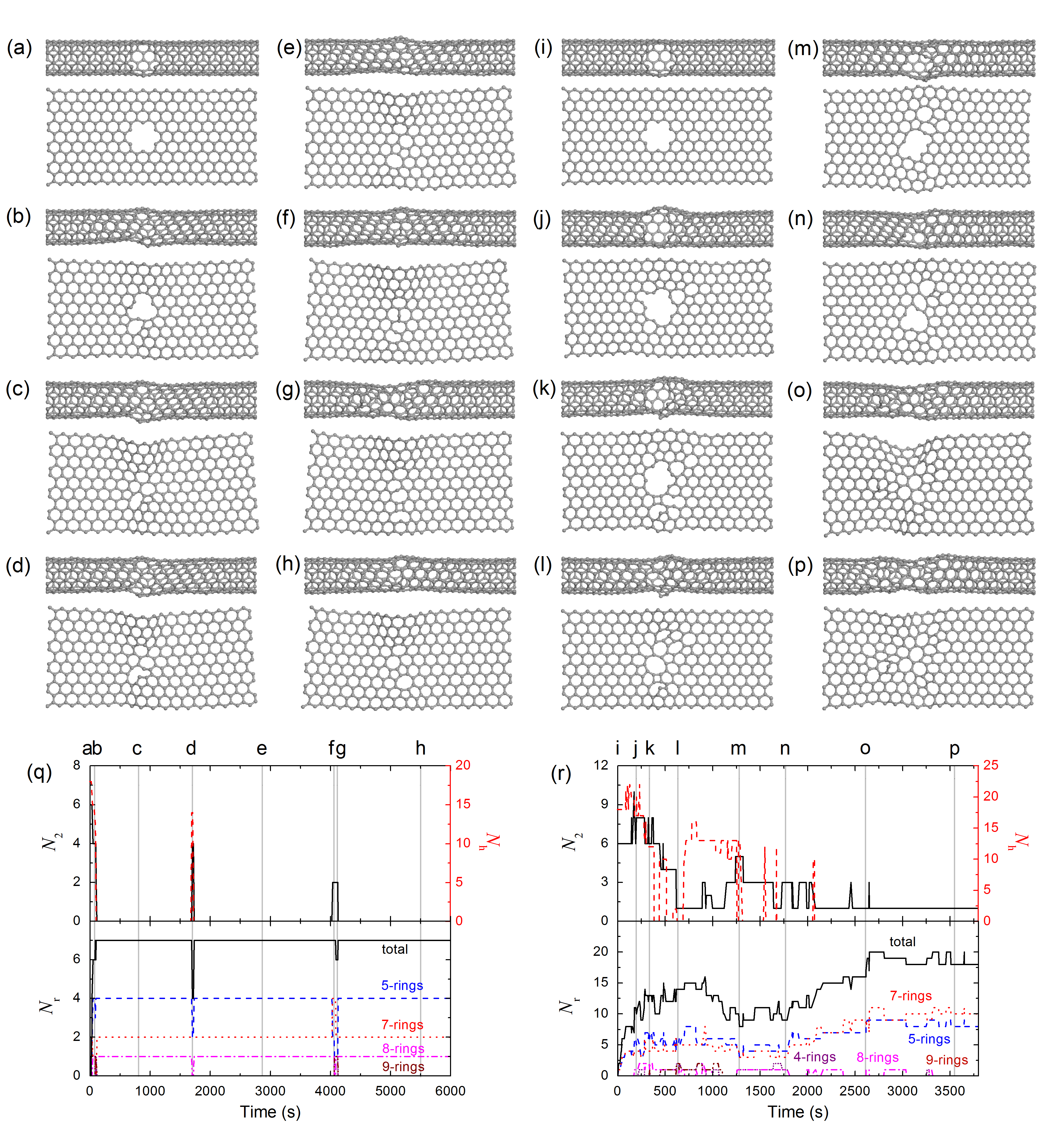}
   \caption{Structure evolution of the (5,5) CNT with a hole under electron irradiation in HRTEM in two simulation runs. In run I (a--h,q), there were very few structural changes after the initial hole healing, while in run II (i--p,r), structural changes occurred continuously. The atomistic structures shown correspond to the moments of time (a) 0 s, (b) 82 s, (c) 816 s, (d) 1698 s, (e) 2871 s, (f) 4061 s, (g) 4111 s and (h) 5494 s for run I and (i) 0 s, (j) 199 s, (k) 336 s, (l) 636 s, (m) 1282 s, (n) 1767 s, (o) 2614 s and (p) 3549 s for run II. To facilitate visualization of structural defects, the structures are unfolded into graphene nanoribbons by cutting  the CNT wall along the axis at the side opposite to the hole location (bottom of each panel). In panels (q) and (r) for runs I and II, respectively, quantities characterizing the structural defects are shown as functions of time (in s): the number $N_2$ of two- and one-coordinated atoms  (top, left axis, black solid lines), number of atoms at hole perimeter $N_\mathrm{h}$  (top, right axis, red dashed lines) and the numbers $N_\mathrm{r}$ of non-hexagonal rings (bottom) of different size: 4- (purple short dotted lines), 5- (blue dashed lines), 7- (red dotted lines), 8- (magenta dash-dotted lines), 9-membered (brown dash-dot-dotted lines) and total (black solid lines). The vertical grey lines indicate the moments of time corresponding to structures (a--p). }
   \label{fig:structure}
\end{figure*}

\begin{figure}
   \centering
   \includegraphics[width=0.5\textwidth]{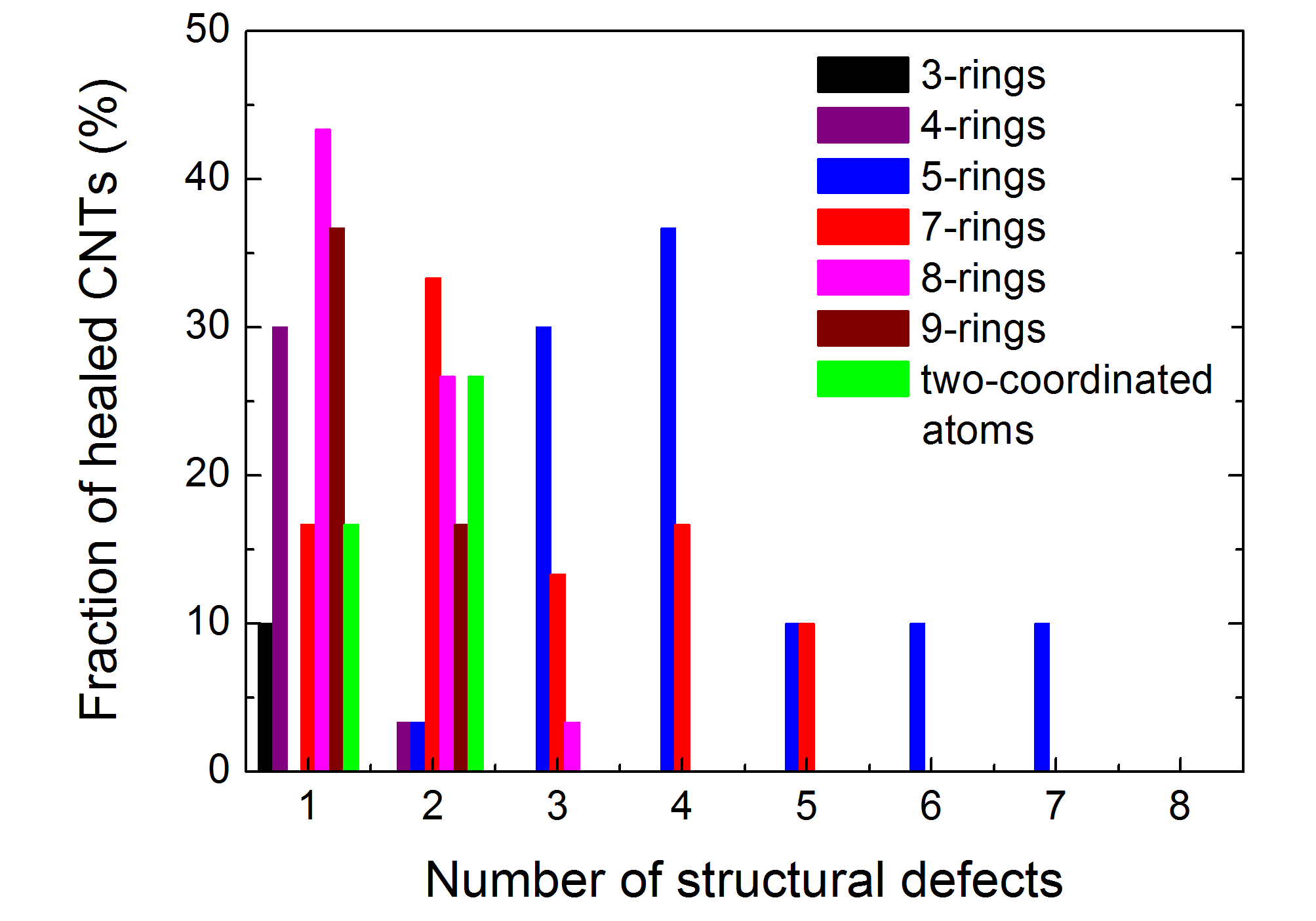}
   \caption{Calculated fractions of CNT structures (in \%) at the moment of the initial hole healing with different numbers of structural defects: 3- (black), 4- (purple), 5- (blue), 7- (red), 8- (magenta) and 9-membered (brown) rings and two-coordinated atoms (green).}
   \label{fig:rings}
\end{figure}

In all 30 simulation runs, the hole was healed and an amorphous patch consisting of non-hexagonal rings  was formed (Figure \ref{fig:structure}). In more than half of the runs, this amorphous patch was growing noticeably under further electron irradiation. The average time required for healing of the initial hole was 390$\pm50$ s. Almost no knock-out of atoms was observed. Only in 5 simulation runs, one atom was ejected before the hole healing. The quality of the CNT structure upon healing can be characterized by the presence of structural defects as compared to the perfect CNT consisting of hexagonal rings only. The distributions of the numbers of rings of different size and the number of two-coordinated atoms in the CNT at this moment are shown in Figure  \ref{fig:rings}. 15 of the corresponding CNT structures involved two-coordinated atoms or short-living 3-membered rings. Among them, 14 CNTs had 3-, 4- and/or 9-membered rings and 1 CNT only 5- to 8-membered rings. 15 structures did not contain one or two-coordinated atoms or 3-membered rings: 5 CNTs included 4- or 9-membered rings, 8 CNTs contained 5- to 8-membered rings, while 2 CNTs consisted of 5- to 7-membered rings only.

Formation of non-hexagonal rings was accompanied by local narrowing of the CNT close to the initial hole (i.e. bottleneck formation) and CNT buckling (Figures \ref{fig:structure}p and 1l, respectively). Formation of such a bottleneck means that the lateral surface of the CNT is decreased. This happens because the lack of carbon atoms that makes it impossible to rebuild the sp$^2$ structure with the same lateral surface as of the pristine CNT. Since healing of the hole in the CNT occurs with shrinkage of the sp$^2$ structure surface, healing of a similar hole in graphene only via structure reconstruction is hardly possible under the same conditions. On the contrary, it has been shown that atomic-scale vacancies in graphene under 80 keV electron irradiation grow into holes of 3--20 \AA~in diameter \cite{Russoat2012,Warner2009}.  

\subsection{Structure Evolution after Hole Healing}

Structure evolution under electron irradiation continued after the initial hole healing. At this stage, two types of CNT structures were observed which can be referred to as stable and unstable ones. In stable structures (Figure \ref{fig:structure}a--h), any structural changes were concentrated in short periods of time with long time intervals between them (Figure \ref{fig:structure}q). Therefore, such events can be considered as rare. In unstable structures, continuous changes and visible amorphous patch growth took place around the former hole after the hole healing  (Figure \ref{fig:structure}i--p,r).

The short periods of structural rearrangements observed in the stable structures and amorphous patch growth in the unstable structures were normally related to the presence of two-coordinated atoms. Occasionally two-coordinated atoms transformed into one-coordinated ones or three-coordinated atoms within 3-membered rings. Such one-coordinated, two-coordinated and three-coordinated atoms within 3-membered rings are referred here to as under-coordinated atoms (UCAs). It was suggested that reactions with participation of two-coordinated atoms play an important role in formation of abundant fullerene isomers like C$_{60}$-I$_\mathrm{h}$ \cite{Sinitsa2020,Eggen1987,Ewels2002}. Namely, autocatalysis by an extra two-coordinated atom in even fullerenes considerably reduces barriers of reactions in which sizes of rings of the sp$^2$ carbon structure (i.e. the structure consisting of  sp$^2$ atoms only which can be reconstructed by removal of two-coordinated atoms and formation of bonds between the atoms connected through two-coordinated atoms) are modified. In our simulations, UCAs also stimulated changes in the topology of the carbon bond network as discussed below. Note that UCAs changed their position across the patch, similar to two-coordinated atoms in fullerenes \cite{Sinitsa2020}.

The stable structures were observed in 19 out of 30 simulation runs. In 5 simulation runs, highly stable CNT structures were formed and no structural changes were detected during the whole run. In 9 simulation runs, rare structural changes occurred but the structure remained stable (Figure \ref{fig:structure}a--h,q). In 5 more simulation runs, the structure changed from stable to unstable and/or vice versa. 

In the stable structures, pairs of short-living UCAs were formed in 24 out of 26 irradiation-induced reactions detected in the structures without UCAs initially. Only in 2 reactions, the rings were rearranged without formation of UCAs. The average interval between the short periods of structural rearrangements in the stable CNT structures exceeded 2100 s. The average duration of the periods of structural rearrangements, i.e. the lifetime of short-living UCAs, was $37\pm7$ s. During this time, one-three reactions with participation of UCAs occurred with the average time between them of $25\pm4$ s. In some cases, formation of UCAs was accompanied by opening of a new hole with 10 or more atoms along the perimeter. The average time required for opening of a new hole exceeded 3300 s, while the average lifetime of new holes was only $17\pm5$ s. 

Short-living UCAs stimulated changes in the topology of the carbon bond network of the stable structures. In a single period of structural rearrangements, the number of non-hexagonal rings could increase, decrease or stay the same (Figure \ref{fig:structure}q). However, there are some trends that become clear from consideration of the estimated average rates at which the numbers of different rings changed during the presence of short-living UCAs. As seen from Table \ref{table:rings}, 4- and 9-membered rings slowly transformed into more stable 5-, 7- and 8-membered rings. The numbers of 5-, 6- and 7-membered rings were changing relatively fast. New 5- and 7-membered rings were formed mostly from 6-membered rings.

\begin{table*}
    \caption{Average rates of changes in the numbers of $n$-membered rings in the stable and unstable CNT structures after the initial hole healing.}
   \renewcommand{\arraystretch}{1.2}
   \setlength{\tabcolsep}{12pt}
    \resizebox{1.0\textwidth}{!}{
        \begin{tabular}{*{9}{c}}
\hline
Structure & Units & \multicolumn{7}{c}{Rings}
 \\\hline
$n$ & &   3 &	4 &	5 &	6 &	7 & 8	& 9
 \\\hline
stable & per pair of UCAs & 0&-0.11 &	0.19 &	-0.58 &	0.42 &	0.05 &	-0.05 \\\hline
unstable & per 1000 s &  -0.060	& -0.040	& 1.080 & -1.921	& 1.080	& -0.020 &	-0.200  \\\hline 
\end{tabular}
}
\label{table:rings}
\end{table*}

Intensive structural reconstruction took place after the initial hole healing in the unstable CNT structures under electron irradiation. In such systems (Figure \ref{fig:structure}i--p), continuous structural changes were also related to the presence of UCAs (Figure \ref{fig:structure}r) but long-living ones. In 15 out of 30 simulation runs, the UCAs were present in the structure of the moment of the initial hole healing. In two of these runs, the UCAs were knocked out. In two other runs, annihilation of the pairs of UCAs \cite{Sinitsa2020} was observed, i.e. UCAs formed a bond thus transforming into three-coordinated atoms. In 3 simulation runs, long-living UCAs were formed after the initial hole healing. In one of these cases, an atom was ejected from a stable CNT structure. In two other cases, a pair of UCAs was formed because  a bond  between two three-coordinated atoms got broken and then the UCAs formed got spatially separated or one of them knocked out.

The average interval between reactions detected in the unstable CNT structures was only 56$\pm$3 s, which is at least 30 times smaller than the interval between the short periods of structural rearrangements in the stable structures but twice greater than the interval between the reactions in the presence of short-living UCAs in the stable structures. Reiterated hole opening and healing was also typical for the unstable CNT structures. 
The average time interval between the events of formation of a new hole was $360\pm60$ s, i.e. at least an order of magnitude smaller than in the stable structures. At the same time, the average time period during which a new hole was observed reached $56\pm14$ s, i.e. 3 times larger than in the stable structures.

During the reactions taking place in  the unstable CNT structures, the number of non-hexagonal rings tended to grow, though there were also periods of descent or stabilization (Figure \ref{fig:structure}r). As seen from Table \ref{table:rings}, 4-, 8- and 9-membered rings slowly transformed into 5- and 7-membered rings. The numbers of the latter ones were growing fast, predominantly via transformations of 6-membered rings. Approximately one pair of 5- and 7-membered rings  was generated each 1000 s. Therefore, both in the stable and unstable CNT structures, selection of 5- to 7-membered rings (more stable than 4-, 8- and 9-membered rings) was observed. Also in the both cases, the growth of the amorphous patch took place. However, the average growth rate was an order of magnitude greater in the unstable structures.

\subsection{Yields of Different Structures}

\begin{figure}
   \centering
   \includegraphics[width=0.5\textwidth]{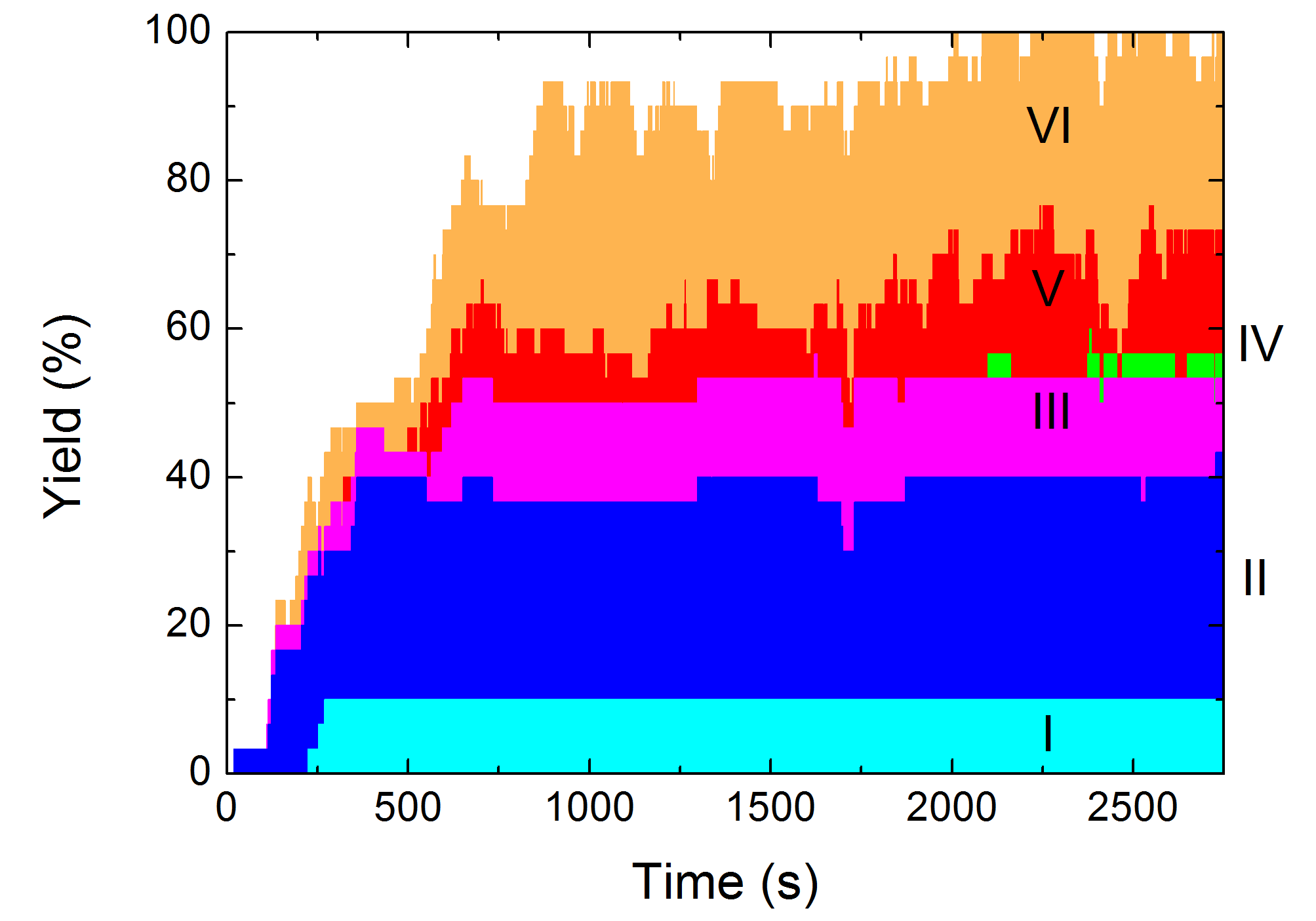}
   \caption{Computed yields (in \%) of CNTs with different structural defects formed upon healing of the (5,5) CNT with a hole under electron irradiation in HRTEM as functions of time (in s). The areas of different colour correspond to the following structures of healed CNTs:  (I) consisting of 5-, 6- and 7-membered rings only, (II) with 8-membered rings, (III) with 4- or 9-membered rings, (IV) with under-coordinated  atoms (UCAs) and consisting of 5-, 6- and 7-membered rings only, (V) with UCAs and 8-membered rings and (VI) with UCAs and 3-, 4- or 9-membered rings. }
   \label{fig:yield}
\end{figure}

Based on the data obtained in the MD simulations, we computed the yields of CNTs with structure of different quality depending on time. As can be appreciated from Figure \ref{fig:yield}, the stable CNT structures, especially without 4- and 9-membered rings, were characterized by shorter healing times (Figure \ref{fig:yield}, on average $270\pm50$ s for the stable structures vs. $510\pm80$ s for the unstable ones). The yields of such structures also changed slowly with time. In 3 out of 30 runs, very high quality structures with only  5- and 7-membered rings among non-hexagonal rings and no UCAs were formed at times 220 -- 270 s and virtually no reactions were detected afterwards. In one
more run, formation of the structure consisting of only these types of rings occurred at 4186 s after significant structural reconstruction.

The yield of the unstable CNTs with UCAs oscillated greatly in time (Figure \ref{fig:yield}) because of the reiterated hole formation. As a result, the $100\%$ yield of healed CNTs was reached only at 2000 s (although the inital hole was healed at times of 20 -- 1200 s). In addition to UCAs, most of the unstable structures contained other defects like 3-, 4- or 9-membered rings. However, the yields of the unstable structures with the rings from 5 to 8 and from 5 to 7  increased with time. 

\subsection{Atomistic Mechanisms of Structural Rearrangements}

\begin{figure*}
   \centering
   \includegraphics[width=\textwidth]{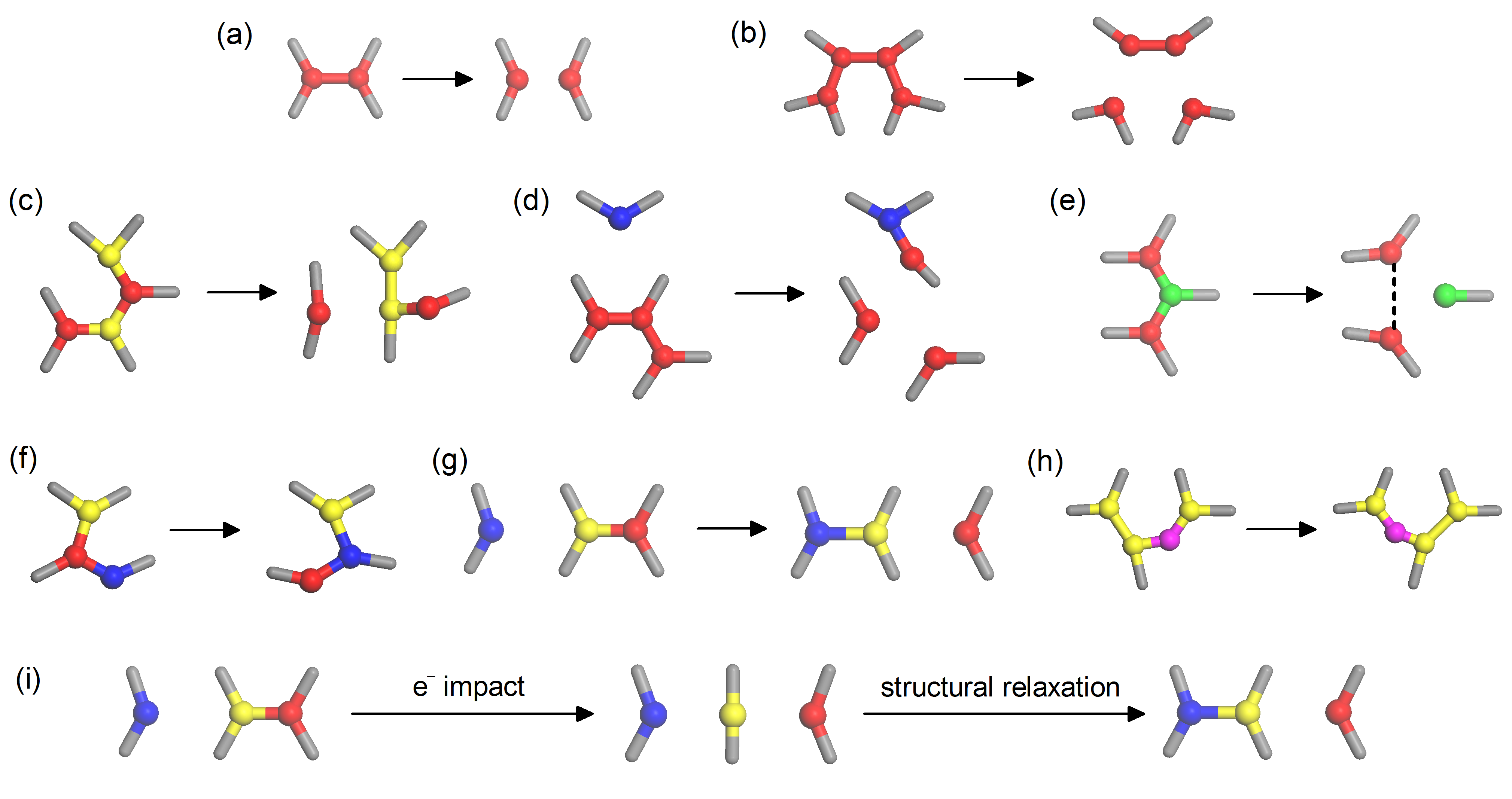}
   \caption{Schemes of reactions induced by electron impacts in the CNT after the initial hole healing: (a) bond breakage between two three-coordinated atoms leading to formation of a pair of two-coordinated atoms, (b) simultaneous breakage of two bonds leading to formation of two pairs of two-coordinated atoms, (c) formation of a pair of two-coordinated atoms and simultaneous migration of one of them (see also Figure 5a), (d) formation of a pair of two-coordinated atoms and simultaneous migration of another two-coordinated atom located nearby (Figure 5b), (e) formation of a one-coordinated atom, (f) short-range sp-defect migration (Figure 5c), (g) long-range sp-defect migration (Figures 5d and e), (h) hopping of a two-coordinated atom, (i) two-step sp-defect migration consisting of bond breakage upon an electron impact followed by bond formation upon the further structural relaxation at longer times. Atoms that become two-coordinated or one-coordinated after the reactions are shown in red and green, respectively. Atoms that are two-coordinated before the reactions are coloured in blue. Three-coordinated atoms that change their neighbours are shown in yellow. The hopping two-coordinated atom in panel (h) is coloured in magenta. In panel (e), the bond that is formed in about 2/3 of the cases is indicated using the dashed line. }
   \label{fig:react}
\end{figure*}

\begin{figure}
   \centering
   \includegraphics[width=0.5\textwidth]{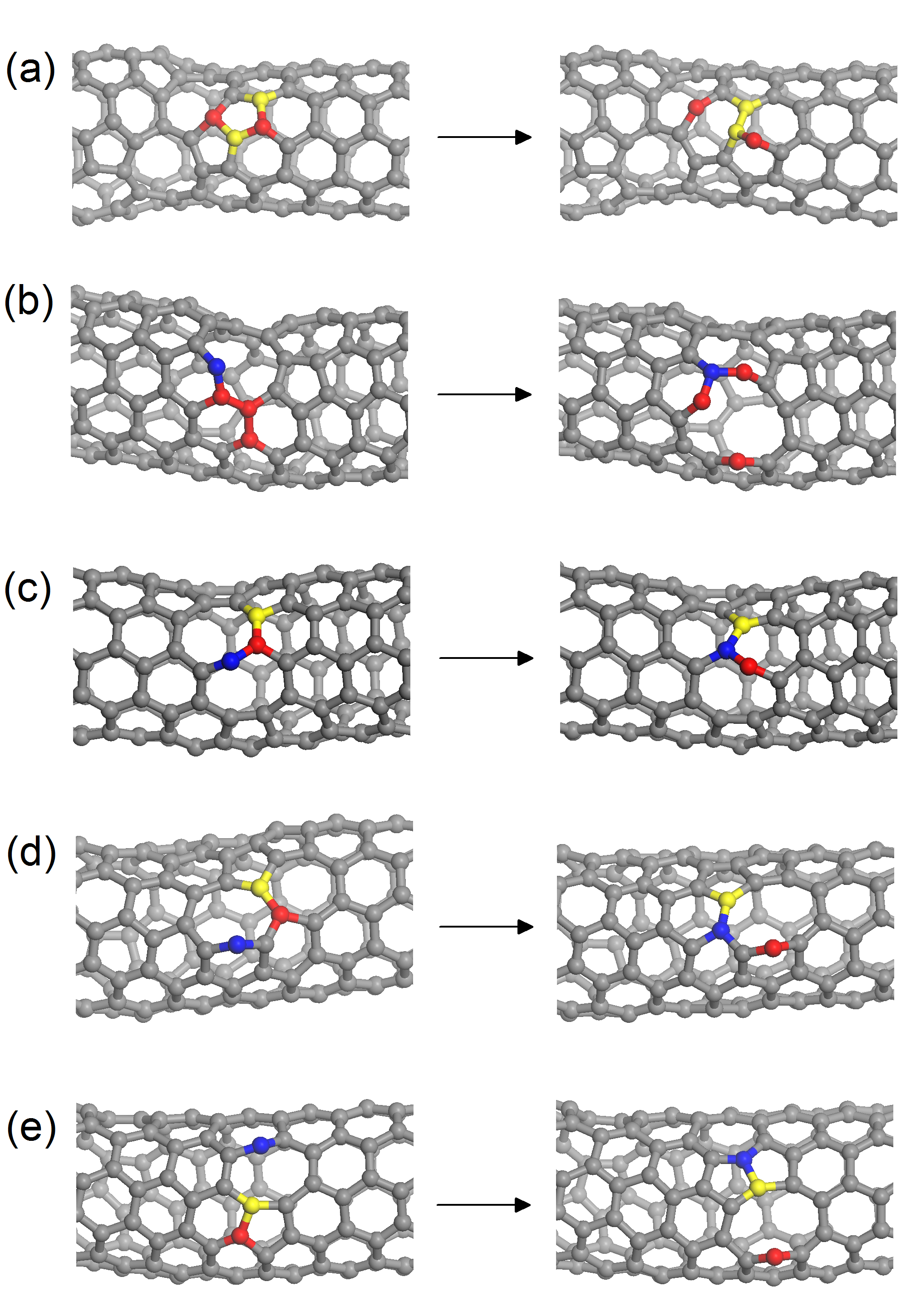}
   \caption{Examples of reactions induced by electron impacts in the CNT after the initial hole healing: (a) formation of a pair of two-coordinated atoms and simultaneous migration of one of them (see also Figure 4c), (b) formation of a pair of two-coordinated atoms and simultaneous migration of another two-coordinated atom located nearby (Figure 4d), (c) short-range sp-defect migration (Figure 4f) and (d,e) long-range sp-defect migration (Figure 4g). Atoms that become two-coordinated after the reactions or are two-coordinated before the reactions are shown in red and blue, respectively. Three-coordinated atoms that change their neighbours are coloured in yellow. }
   \label{fig:rexample}
\end{figure}

Let us now discuss the atomistic mechanisms of structural rearrangements under electron irradiation after the initial hole healing. General schemes of bond rearrangement reactions induced in the CNT by electron impacts are given in Figure \ref{fig:react}.  The small fragments that were used to differentiate the reactions in the numerical analysis are shown. The local structure around the atoms participating the reactions was very diverse in different detected cases of the same reaction. Examples of some reactions including the local CNT structure are shown in Figure \ref{fig:rexample}. 

Formation of pairs of UCAs occurred in the following processes: breakage of a  bond between two three-coordinated atoms (Figure \ref{fig:react}a), simultaneous breakage of two bonds between adjacent pairs of atoms (Figure \ref{fig:react}b), breakage of a bond and simultaneous migration of one of the two-coordinated atoms formed (Figure \ref{fig:react}c and Figure \ref{fig:rexample}a),  formation of a pair of two-coordinated atoms and simultaneous migration of another two-coordinated atom located nearby (Figure \ref{fig:react}d and Figure \ref{fig:rexample}b), simultaneous breakage of two bonds of a three-coordinated atom leading to formation of a one-coordinated atom and a pair of two-coordinated atoms (Figure \ref{fig:react}e), etc. 

Two-coordinated atoms moved easily across the amorphous patch according to the following mechanism (Figures \ref{fig:react}f,g and Figures \ref{fig:rexample}c--e): a bond was broken nearby the two-coordinated atom between two three-coordinated atoms  and one of them formed a new bond with the two-coordinated atom. This can be considered as generalization of the mechanism of sp-defect migration  \cite{Sinitsa2020} revealed previously in fullerenes with an odd number of atoms. The term of ``sp-defect migration" was  introduced initially for the case when the new and former two-coordinated atoms are bonded. In this case (which we refer to here as ``short-range sp-defect migration", Figure \ref{fig:react}f), the process simply leads to displacement of the two-coordinated atom and no change in rings of the sp$^2$ structure occurs (Figure \ref{fig:rexample}c). However, if the new and former two-coordinated atoms are not bonded (``long-range sp-defect migration", Figure \ref{fig:react}g), the process results both in displacement of the two-coordinated atom and changes in rings of the sp$^2$ structure (Figures \ref{fig:rexample}d and e), i.e. corresponds to an autocatalytic reaction. Note that one of the reaction pathways for autocatalyzed Stone-Wales rearrangement proposed previously \cite{Ewels2002} complies with the mechanism of long-range sp-defect migration. An example of the corresponding reaction detected in our simulations is shown in Figure \ref{fig:rexample}d. In addition to the migration events discussed, several examples of hopping of the same two-coordinated atom to another place (Figure \ref{fig:react}h) were observed. 

Structural relaxation between electron impacts was also important for the amorphous patch growth. Annihilation of pairs of two-coordinated atoms, the reaction reverse to bond breakage shown in Figure \ref{fig:react}a, happened only at the high-temperature stage of the MD simulations during the structural relaxation between electron impacts. Multistep processes including bond breakage upon an electron impact leading to formation of UCAs and bond formation upon structural relaxation between electron impacts resulting in annihilation of pairs of UCAs were detected as well. In particular, such multistep processes contributed to migration of two-coordinated atoms and changes in sizes of rings of the sp$^2$ structure. For example, two-step short-range and long-range (Figure \ref{fig:react}i) sp-defect migration events were observed when a bond was broken in the vicinity of a pre-existing two-coordinated atom. 

The frequencies of different irradiation-induced events in the MD simulations strongly depended on the number of bonds broken. Almost all (96\%) successful electron impacts, i.e. those producing any structural change, induced bond breakage between two three-coordinated atoms (Figure \ref{fig:react}a). However, normally this bond was restored during the structural relaxation stage of the MD simulations and we do not discuss these cases. Nevertheless, among the reactions that were not reversed before the next electron impact (877 of such reactions in total for all the simulation runs), the great majority still corresponded to those where only one bond was broken. It was bond breakage between two three-coordinated atoms with formation of a pair of two-coodinated atoms (Figure \ref{fig:react}a) in 48\% of the reactions and generalized sp-defect migration (Figures \ref{fig:react}f,g and Figures \ref{fig:rexample}c--e) in 42\%. In the rest of the reactions two bonds were broken simultaneously (Figure \ref{fig:react}b-e and h). Only one event with simultaneous breakage of three bonds was observed and it was knock-out of a three-coordinated atom from a 5-membered ring. Thus, there was a correlation between the number of broken bonds and the probability of the reaction to take place under electron irradiation. The larger number of the bonds was broken, the lower was the probability. This is similar to the case of thermally activated processes, where the probability is determined by the barrier, which grows with the number of bonds broken simultaneously  \cite{Lebedeva2008}.

Long-range sp-defect migration (Figure \ref{fig:react}g and Figures \ref{fig:rexample}d and e) occured more frequently than short-range one (Figure \ref{fig:react}f and Figure \ref{fig:rexample}c), 24.6\% versus 16.9\% of the reactions, respectively. The two-step mechanism involving structure relaxation at longer times (Figure \ref{fig:react}i) further increased the fractions of long and short-range migration events by  28.9\% and 8.1\%, respectively. For comparison, in studies of structural rearrangements of odd fullerenes at high temperature \cite{Sinitsa2020}, 98\% of the reactions detected corresponded to short-range sp-defect migration.

In our simulations, we found 6 examples of hopping of a two-coordinated atom (Figure \ref{fig:react}h). This reaction was much less frequent than sp-defect migration because it requires simultaneous breakage of two bonds. Note that it was not observed at all in studies of odd fullerenes at high temperature \cite{Sinitsa2020}. 
Other reactions with breakage of two bonds were also rare. Formation of a pair of two-coordinated atoms with simultaneous migration of one of them (Figure \ref{fig:react}c and Figure \ref{fig:rexample}a) or another two-coordinated atom located nearby (Figure \ref{fig:react}d and Figure \ref{fig:rexample}b) took place only 24 and 7 times, i.e. in 2.7\% and 0.8\% of the reactions, respectively. Formation of one-coordinated atoms according to the mechanism from Figure \ref{fig:react}e and breakage of two bonds between several neighbour three-coordinated atoms similar to the one shown in Figure \ref{fig:react}b occured 15 times each, i.e. in 1.7\% of the reactions. Knock-out of  two-coordinated atoms took place 6 times, i.e. in 0.7\% of the reactions (note that knock-out of one-coordinated atoms was also detected only 7 times but here it was related to the difficulty of observing this type of atoms). 

\section{Discussion and Conclusions}
Thus, our MD simulations showed that healing of a six-atom hole in a CNT can occur through structure reconstruction caused by electron impacts in HRTEM in the absence of a source of carbon adatoms. Previously hole healing  was demonstrated only through filling by carbon adatoms \cite{Zan2012,Xu2012,Botari2016}.
Note that adatom motion in the latter case requires high temperature \cite{Song2011,Chen2013,Kotakoski2007,Botari2016}, while the process studied here takes place at room temperature analogously to structure reconstruction of CNTs \cite{Ajayan1998} and graphene \cite{Robertson2012,Kotakoski2011} via motion and coalescence of monovacancies. 

Because of the lack of carbon atoms in our simulations, the perfect CNT structure could not be formed and the patch generated upon the hole healing was amorphous. 
The structures obtained here are consistent with the structures of irradiated graphene and CNTs from literature. 
Various defects present in our CNT structures, such as UCAs, 4-, 5-, 7- and 8-membered rings, were created and observed using HRTEM in graphene \cite{Robertson2012,Robertson2013,Kotakoski2011}. Similar to our results, previous experimental \cite{Robertson2012,Kotakoski2011} and theoretical \cite{Ajayan1998} works indicated that too small or big rings tend to transform into the rings of size from 5 to 7 upon increasing the exposure time. Electron irradiation of graphene in  HRTEM with the energy of electrons slightly below the threshold for knock-on damage of the pristine structure  leads to migration and reconstruction of pre-existing vacancies \cite{Robertson2012,Robertson2013,Kotakoski2011} or growth of such vacancies into holes \cite{Russoat2012,Warner2009}. Generation of vacancies of graphene can be also achieved at a large radiation dose \cite{Robertson2012,Robertson2013}. However, it occurs at random places thoughout the exposed area. The continuous growth of the amorphous patch around the initial seed taking place in our simulations under electron irradiation has not been demonstrated before. 

We found out that the atomistic mechanism of the amorphous patch growth was related to presence of two-coordinated atoms in the patch structure. Such atoms promoted reactions leading to formation of pentagons and heptagons from the initial hexagon structure of the CNT in the way analogous to autocatalytic reactions in fullerenes which result in selection of abundant fullerene isomers \cite{Sinitsa2020,Eggen1987,Ewels2002}. The principal irradiation-induced events in the presence of two-coordinated atoms could be described as generalized sp-defect migration \cite{Sinitsa2020}: a bond was broken between two three-coordinated atoms and one of them formed a new bond with a nearby two-coordinated atom. If the new and former two-coordinated atoms were not bonded, the process led both to displacement of the two-coordinated atom and changes in rings of the sp$^2$ structure. Hopping of the same two-coordinated atom to neighbour position (like adatom migration on perfect hexagon surface of graphene or a nanotube) was also observed but much rarely because it requires breakage of two bonds simultaneously. Sp-defect migration occurred directly as a result of an electron impact only or in two steps with bond breakage upon an electron impact and bond formation during the subsequent structural relaxation at longer times.

Along with formation of the amorphous patch, our simulations revealed a local reduction of the CNT diameter i.e. formation of the bottleneck of about 1 nm in length. Previously  considerable local shrinkage of multi-walled CNTs due to knock-out of atoms with subsequent diffusion of interstitials inside the CNTs was observed under electron irradiation with the high energy of 300 keV \cite{Banhart2005,Sun2006}. In that case, the length of the amorphous bottleneck was considerably longer and equal to the diameter of the electron beam of about 15 nm \cite{Banhart2005,Sun2006}. Bottleneck formation was predicted also by atomistic simulations of plastic deformation of CNTs through thermally activated glide of pentagon-heptagon pairs and C$_2$ molecule emission \cite{Ding2007,Ding2007a}. However, the local diameter decrease at the scale of about 1 nm was observed only for very short CNTs of about 10 nm length \cite{Asaka2005}. Plastic deformation of long CNTs resulted in the uniform diameter decrease along the CNT\cite{Huang2005,Huang2006}. In early 1990s (see, for example, Ref. \onlinecite{Lyo1991}), the methods of controlled formation of holes of a nanometer size using a tip of scanning tunneling microscope were developed. Such methods can be used to create holes on the lateral surface of SWCNTs at predetermined places for subsequent transformation into amorphous patches and bottleneck formation under electron irradiation in HRTEM. 

Let us now discuss possible applications of CNTs with a bottleneck. Significant local shrinkage of multi-walled CNTs causes deformation, extrusion and breaking of the solid material inside the CNTs \cite{Sun2006}. The small local diameter decrease found here gives rise to confined electronic states, which can be used to create CNT-based quantum dots \cite{Chico1998,Liu2003,Ayuela2008} and ambipolar field effect transistors \cite{Triozon2005}. The possibility to grow the amorphous structure region around the patch, as observed in our simulation runs, provides additional means to control the patch size and CNT geometry and thus CNT electronic properties. Methods of SWCNT filling by different materials have been elaborated recently (see, for example, Ref. \onlinecite{Sloan2002} for a review). Filling of SWCNT up to the bottleneck gives another degree of freedom in design of devices with half-filled SWCNTs.

\section*{Ackowledgements}
AMP and AAK acknowledge the Russian Foundation of Basic Research (Grant 18-02-00985). This work has been carried out using computing resources of the federal collective usage center Complex for Simulation and Data Processing for Mega-science Facilities at NRC ``Kurchatov Institute", http://ckp.nrcki.ru/.

\section*{Data availability}
The data that support the findings of this study are openly available in Mendeley Data at http://dx.doi.org/10.17632/wpwjtxxbz5.1 (Lebedeva, 2020).

\section*{References}
\bibliography{jpcc}

\providecommand{\noopsort}[1]{}\providecommand{\singleletter}[1]{#1}%
\providecommand{\latin}[1]{#1}
\providecommand*\mcitethebibliography{\thebibliography}
\csname @ifundefined\endcsname{endmcitethebibliography}
  {\let\endmcitethebibliography\endthebibliography}{}
\begin{mcitethebibliography}{50}
\providecommand*\natexlab[1]{#1}
\providecommand*\mciteSetBstSublistMode[1]{}
\providecommand*\mciteSetBstMaxWidthForm[2]{}
\providecommand*\mciteBstWouldAddEndPuncttrue
  {\def\EndOfBibitem{\unskip.}}
\providecommand*\mciteBstWouldAddEndPunctfalse
  {\let\EndOfBibitem\relax}
\providecommand*\mciteSetBstMidEndSepPunct[3]{}
\providecommand*\mciteSetBstSublistLabelBeginEnd[3]{}
\providecommand*\EndOfBibitem{}
\mciteSetBstSublistMode{f}
\mciteSetBstMaxWidthForm{subitem}{(\alph{mcitesubitemcount})}
\mciteSetBstSublistLabelBeginEnd
  {\mcitemaxwidthsubitemform\space}
  {\relax}
  {\relax}

\bibitem[Song \latin{et~al.}(2011)Song, Schneider, Xu, Pandraud, Dekker, and
  Zandbergen]{Song2011}
Song,~B.; Schneider,~G.~F.; Xu,~Q.; Pandraud,~G.; Dekker,~C.; Zandbergen,~H.
  Atomic-Scale Electron-Beam Sculpting of Near-Defect-Free Graphene
  Nanostructures. \emph{Nano Lett.} \textbf{2011}, \emph{11}, 2247--2250\relax
\mciteBstWouldAddEndPuncttrue
\mciteSetBstMidEndSepPunct{\mcitedefaultmidpunct}
{\mcitedefaultendpunct}{\mcitedefaultseppunct}\relax
\EndOfBibitem
\bibitem[Zan \latin{et~al.}(2012)Zan, Ramasse, Bangert, and Novoselov]{Zan2012}
Zan,~R.; Ramasse,~Q.~M.; Bangert,~U.; Novoselov,~K.~S. Graphene Reknits Its
  Holes. \emph{Nano Lett.} \textbf{2012}, \emph{12}, 3936--3940\relax
\mciteBstWouldAddEndPuncttrue
\mciteSetBstMidEndSepPunct{\mcitedefaultmidpunct}
{\mcitedefaultendpunct}{\mcitedefaultseppunct}\relax
\EndOfBibitem
\bibitem[Xu \latin{et~al.}(2012)Xu, Yin, Xie, He, Wang, and Sun]{Xu2012}
Xu,~T.; Yin,~K.; Xie,~X.; He,~L.; Wang,~B.; Sun,~L. Size-Dependent Evolution of
  Graphene Nanopores under Thermal Excitation. \emph{Small} \textbf{2012},
  \emph{8}, 3422--3426\relax
\mciteBstWouldAddEndPuncttrue
\mciteSetBstMidEndSepPunct{\mcitedefaultmidpunct}
{\mcitedefaultendpunct}{\mcitedefaultseppunct}\relax
\EndOfBibitem
\bibitem[Robertson \latin{et~al.}(2014)Robertson, Lee, He, Yoon, Kirkland, and
  Warner]{Robertson2013}
Robertson,~A.~W.; Lee,~G.-D.; He,~K.; Yoon,~E.; Kirkland,~A.~I.; Warner,~J.~H.
  The Role of the Bridging Atom in Stabilizing Odd Numbered Graphene Vacancies.
  \emph{Nano Lett.} \textbf{2014}, \emph{14}, 3972--3980\relax
\mciteBstWouldAddEndPuncttrue
\mciteSetBstMidEndSepPunct{\mcitedefaultmidpunct}
{\mcitedefaultendpunct}{\mcitedefaultseppunct}\relax
\EndOfBibitem
\bibitem[Chen \latin{et~al.}(2013)Chen, Shi, Cai, Xu, Sun, Wu, and
  Yu]{Chen2013}
Chen,~J.; Shi,~T.; Cai,~T.; Xu,~T.; Sun,~L.; Wu,~X.; Yu,~D. Self Healing of
  Defected Gaphene. \emph{Appl. Phys. Lett.} \textbf{2013}, \emph{102},
  103107\relax
\mciteBstWouldAddEndPuncttrue
\mciteSetBstMidEndSepPunct{\mcitedefaultmidpunct}
{\mcitedefaultendpunct}{\mcitedefaultseppunct}\relax
\EndOfBibitem
\bibitem[Ajayan \latin{et~al.}(1998)Ajayan, Ravikumar, and
  Charlier]{Ajayan1998}
Ajayan,~P.~M.; Ravikumar,~V.; Charlier,~J.-C. Surface Reconstructions and
  Dimensional Changes in Single-Walled Carbon Nanotubes. \emph{Phys. Rev.
  Lett.} \textbf{1998}, \emph{81}, 1437--1440\relax
\mciteBstWouldAddEndPuncttrue
\mciteSetBstMidEndSepPunct{\mcitedefaultmidpunct}
{\mcitedefaultendpunct}{\mcitedefaultseppunct}\relax
\EndOfBibitem
\bibitem[Kotakoski \latin{et~al.}(2011)Kotakoski, Krasheninnikov, Kaiser, and
  Meyer]{Kotakoski2011}
Kotakoski,~J.; Krasheninnikov,~A.~V.; Kaiser,~U.; Meyer,~J.~C. From Point
  Defects in Graphene to Two-Dimensional Amorphous Carbon. \emph{Phys. Rev.
  Lett.} \textbf{2011}, \emph{106}, 105505\relax
\mciteBstWouldAddEndPuncttrue
\mciteSetBstMidEndSepPunct{\mcitedefaultmidpunct}
{\mcitedefaultendpunct}{\mcitedefaultseppunct}\relax
\EndOfBibitem
\bibitem[Robertson \latin{et~al.}(2012)Robertson, Allen, Wu, He, Olivier,
  Neethling, Kirkland, and Warner]{Robertson2012}
Robertson,~A.~W.; Allen,~C.~S.; Wu,~Y.~A.; He,~K.; Olivier,~J.; Neethling,~J.;
  Kirkland,~A.~I.; Warner,~J.~H. Spatial Control of Defect Creation in Graphene
  at the Nanoscale. \emph{Nature Comm.} \textbf{2012}, \emph{3}, 1144\relax
\mciteBstWouldAddEndPuncttrue
\mciteSetBstMidEndSepPunct{\mcitedefaultmidpunct}
{\mcitedefaultendpunct}{\mcitedefaultseppunct}\relax
\EndOfBibitem
\bibitem[Girit \latin{et~al.}(2009)Girit, Meyer, Erni, Rossell, Kisielowski,
  Yang, Park, Crommie, Cohen, Louie, and Zettl]{Girit2009}
Girit,~{\c C}.~{\"O}.; Meyer,~J.~C.; Erni,~R.; Rossell,~M.~D.; Kisielowski,~C.;
  Yang,~L.; Park,~C.-H.; Crommie,~M.~F.; Cohen,~M.~L.; Louie,~S.~G.; Zettl,~A.
  Graphene at the Edge: Stability and Dynamics. \emph{Science} \textbf{2009},
  \emph{323}, 1705--1708\relax
\mciteBstWouldAddEndPuncttrue
\mciteSetBstMidEndSepPunct{\mcitedefaultmidpunct}
{\mcitedefaultendpunct}{\mcitedefaultseppunct}\relax
\EndOfBibitem
\bibitem[Krasheninnikov \latin{et~al.}(2002)Krasheninnikov, Nordlund, and
  Keinonen]{Krasheninnikov2002}
Krasheninnikov,~A.~V.; Nordlund,~K.; Keinonen,~J. Production of Defects in
  Supported Carbon Nanotubes under Ion Irradiation. \emph{Phys. Rev. B}
  \textbf{2002}, \emph{65}, 165423\relax
\mciteBstWouldAddEndPuncttrue
\mciteSetBstMidEndSepPunct{\mcitedefaultmidpunct}
{\mcitedefaultendpunct}{\mcitedefaultseppunct}\relax
\EndOfBibitem
\bibitem[Jiang and Wang(2010)Jiang, and Wang]{Jiang2010}
Jiang,~J.-W.; Wang,~J.-S. Self-Repairing in Single-Walled Carbon Nanotubes by
  Heat Treatment. \emph{J. Appl. Phys.} \textbf{2010}, \emph{108}, 054503\relax
\mciteBstWouldAddEndPuncttrue
\mciteSetBstMidEndSepPunct{\mcitedefaultmidpunct}
{\mcitedefaultendpunct}{\mcitedefaultseppunct}\relax
\EndOfBibitem
\bibitem[Kotakoski \latin{et~al.}(2007)Kotakoski, Krasheninnikov, and
  Nordlund]{Kotakoski2007}
Kotakoski,~J.; Krasheninnikov,~A.~V.; Nordlund,~K. Kinetic Monte Carlo
  Simulations of the Response of Carbon Nanotubes to Electron Irradiation.
  \emph{J. of Computational and Theoretical Nanoscience} \textbf{2007},
  \emph{4}, 1153--1159\relax
\mciteBstWouldAddEndPuncttrue
\mciteSetBstMidEndSepPunct{\mcitedefaultmidpunct}
{\mcitedefaultendpunct}{\mcitedefaultseppunct}\relax
\EndOfBibitem
\bibitem[Botari \latin{et~al.}(2016)Botari, Paupitz, {Alves da Silva Autreto},
  and Galvao]{Botari2016}
Botari,~T.; Paupitz,~R.; {Alves da Silva Autreto},~P.; Galvao,~D.~S. Graphene
  Healing Mechanisms: A Theoretical Investigation. \emph{Carbon} \textbf{2016},
  \emph{99}, 302--309\relax
\mciteBstWouldAddEndPuncttrue
\mciteSetBstMidEndSepPunct{\mcitedefaultmidpunct}
{\mcitedefaultendpunct}{\mcitedefaultseppunct}\relax
\EndOfBibitem
\bibitem[Lee \latin{et~al.}(2014)Lee, Ryu, Lee, and Chang]{Lee2014}
Lee,~A.~T.; Ryu,~B.; Lee,~I.-H.; Chang,~K.~J. Action-Derived Molecular Dynamics
  Simulations for the Migration and Coalescence of Vacancies in Graphene and
  Carbon Nanotubes. \emph{J. Phys.: Condens. Matter} \textbf{2014}, \emph{26},
  115303\relax
\mciteBstWouldAddEndPuncttrue
\mciteSetBstMidEndSepPunct{\mcitedefaultmidpunct}
{\mcitedefaultendpunct}{\mcitedefaultseppunct}\relax
\EndOfBibitem
\bibitem[Trevethan \latin{et~al.}(2014)Trevethan, Latham, Heggie, Briddon, and
  Rayson]{Trevethan2014}
Trevethan,~T.; Latham,~C.~D.; Heggie,~M.~I.; Briddon,~P.~R.; Rayson,~M.~J.
  Vacancy Diffusion and Coalescence in Graphene Directed by Defect Strain
  Fields. \emph{Nanoscale} \textbf{2014}, \emph{6}, 2978--2986\relax
\mciteBstWouldAddEndPuncttrue
\mciteSetBstMidEndSepPunct{\mcitedefaultmidpunct}
{\mcitedefaultendpunct}{\mcitedefaultseppunct}\relax
\EndOfBibitem
\bibitem[Santana \latin{et~al.}(2013)Santana, Popov, and
  Bichoutskaia]{Santana2013a}
Santana,~A.; Popov,~A.~M.; Bichoutskaia,~E. Stability and Dynamics of Vacancy
  in Graphene Flakes: Edge Effects. \emph{Chem. Phys. Lett.} \textbf{2013},
  \emph{557}, 80--87\relax
\mciteBstWouldAddEndPuncttrue
\mciteSetBstMidEndSepPunct{\mcitedefaultmidpunct}
{\mcitedefaultendpunct}{\mcitedefaultseppunct}\relax
\EndOfBibitem
\bibitem[Van \latin{et~al.}(2017)Van, Badura, Liang, Okoli, and Zhang]{Van2017}
Van,~H.~H.; Badura,~K.; Liang,~R.; Okoli,~O.; Zhang,~M. Laser-Induced Graphitic
  Healing of Carbon Nanotubes Aligned in a Sheet. \emph{J. Laser Appl.}
  \textbf{2017}, \emph{29}, 022010\relax
\mciteBstWouldAddEndPuncttrue
\mciteSetBstMidEndSepPunct{\mcitedefaultmidpunct}
{\mcitedefaultendpunct}{\mcitedefaultseppunct}\relax
\EndOfBibitem
\bibitem[Maheshwari \latin{et~al.}(2012)Maheshwari, Singh, and
  Mathur]{Maheshwari2012}
Maheshwari,~P.~H.; Singh,~R.; Mathur,~R.~B. Effect of Heat Treatment on the
  Structure and Stability of Multiwalled Carbon Nanotubes Produced by Catalytic
  Chemical Vapor Deposition Technique. \emph{Mater. Chem. Phys.} \textbf{2012},
  \emph{134}, 412--416\relax
\mciteBstWouldAddEndPuncttrue
\mciteSetBstMidEndSepPunct{\mcitedefaultmidpunct}
{\mcitedefaultendpunct}{\mcitedefaultseppunct}\relax
\EndOfBibitem
\bibitem[Pierlot \latin{et~al.}(2014)Pierlot, Woodhead, and
  Church]{Pierlot2014}
Pierlot,~A.~P.; Woodhead,~A.~L.; Church,~J.~S. Thermal Annealing Effects on
  Multi-Walled Carbon Nanotube Yarns Probed by Raman Spectroscopy.
  \emph{Spectrochim. Acta - Part A Mol. Biomol. Spectrosc.} \textbf{2014},
  \emph{117}, 598--603\relax
\mciteBstWouldAddEndPuncttrue
\mciteSetBstMidEndSepPunct{\mcitedefaultmidpunct}
{\mcitedefaultendpunct}{\mcitedefaultseppunct}\relax
\EndOfBibitem
\bibitem[Banhart \latin{et~al.}(2011)Banhart, Kotakoski, and
  Krasheninnikov]{Banhart2011}
Banhart,~F.; Kotakoski,~J.; Krasheninnikov,~A.~V. Structural Defects in
  Graphene. \emph{ACS Nano} \textbf{2011}, \emph{5}, 26--41\relax
\mciteBstWouldAddEndPuncttrue
\mciteSetBstMidEndSepPunct{\mcitedefaultmidpunct}
{\mcitedefaultendpunct}{\mcitedefaultseppunct}\relax
\EndOfBibitem
\bibitem[Santana \latin{et~al.}(2013)Santana, Zobelli, Kotakoski, Chuvilin, and
  Bichoutskaia]{Santana2013}
Santana,~A.; Zobelli,~A.; Kotakoski,~J.; Chuvilin,~A.; Bichoutskaia,~E.
  Inclusion of Radiation Damage Dynamics in High-Resolution Transmission
  Electron Microscopy Image Simulations: The Example of Graphene. \emph{Phys.
  Rev. B} \textbf{2013}, \emph{87}, 094110\relax
\mciteBstWouldAddEndPuncttrue
\mciteSetBstMidEndSepPunct{\mcitedefaultmidpunct}
{\mcitedefaultendpunct}{\mcitedefaultseppunct}\relax
\EndOfBibitem
\bibitem[Skowron \latin{et~al.}(2013)Skowron, Lebedeva, Popov, and
  Bichoutskaia]{Skowron2013}
Skowron,~S.~T.; Lebedeva,~I.~V.; Popov,~A.~M.; Bichoutskaia,~E. Approaches to
  Modelling Irradiation-Induced Processes in Transmission Electron Microscopy.
  \emph{Nanoscale} \textbf{2013}, \emph{5}, 6677--6692\relax
\mciteBstWouldAddEndPuncttrue
\mciteSetBstMidEndSepPunct{\mcitedefaultmidpunct}
{\mcitedefaultendpunct}{\mcitedefaultseppunct}\relax
\EndOfBibitem
\bibitem[Sinitsa \latin{et~al.}(2017)Sinitsa, Chamberlain, Zoberbier, Lebedeva,
  Popov, Knizhnik, McSweeney, Biskupek, Kaiser, and Khlobystov]{Sinitsa2017}
Sinitsa,~A.~S.; Chamberlain,~T.~W.; Zoberbier,~T.; Lebedeva,~I.~V.;
  Popov,~A.~M.; Knizhnik,~A.~A.; McSweeney,~R.~L.; Biskupek,~J.; Kaiser,~U.;
  Khlobystov,~A.~N. Formation of Nickel Clusters Wrapped in Carbon Cages:
  Towards New Endohedral Metallofullerene Synthesis. \emph{Nano Lett.}
  \textbf{2017}, \emph{17}, 1082--1089\relax
\mciteBstWouldAddEndPuncttrue
\mciteSetBstMidEndSepPunct{\mcitedefaultmidpunct}
{\mcitedefaultendpunct}{\mcitedefaultseppunct}\relax
\EndOfBibitem
\bibitem[Lebedeva \latin{et~al.}(2014)Lebedeva, Chamberlain, Popov, Knizhnik,
  Zoberbier, Biskupek, Kaiser, and Khlobystov]{Lebedeva2014}
Lebedeva,~I.~V.; Chamberlain,~T.~W.; Popov,~A.~M.; Knizhnik,~A.~A.;
  Zoberbier,~T.; Biskupek,~J.; Kaiser,~U.; Khlobystov,~A.~N. Atomistic
  Mechanism of Carbon Nanotube Cutting Catalyzed by Nickel under an Electron
  Beam. \emph{Nanoscale} \textbf{2014}, \emph{6}, 14877--14890\relax
\mciteBstWouldAddEndPuncttrue
\mciteSetBstMidEndSepPunct{\mcitedefaultmidpunct}
{\mcitedefaultendpunct}{\mcitedefaultseppunct}\relax
\EndOfBibitem
\bibitem[Knizhnik(2020)]{Knizhnik2017}
Knizhnik,~A.~A. {{\it{MD-kMC}}; Kintech Lab Ltd.: Moscow}, 2020\relax
\mciteBstWouldAddEndPuncttrue
\mciteSetBstMidEndSepPunct{\mcitedefaultmidpunct}
{\mcitedefaultendpunct}{\mcitedefaultseppunct}\relax
\EndOfBibitem
\bibitem[Sinitsa \latin{et~al.}(2018)Sinitsa, Lebedeva, Popov, and
  Knizhnik]{Sinitsa2018}
Sinitsa,~A.~S.; Lebedeva,~I.~V.; Popov,~A.~M.; Knizhnik,~A.~A. Long Triple
  Carbon Chains Formation by Heat Treatment of Graphene Nanoribbon: Molecular
  Dynamics Study with Revised Brenner Potential. \emph{Carbon} \textbf{2018},
  \emph{140}, 543--556\relax
\mciteBstWouldAddEndPuncttrue
\mciteSetBstMidEndSepPunct{\mcitedefaultmidpunct}
{\mcitedefaultendpunct}{\mcitedefaultseppunct}\relax
\EndOfBibitem
\bibitem[Sinitsa \latin{et~al.}(2019)Sinitsa, Lebedeva, Popov, and
  Knizhnik]{Sinitsa2019}
Sinitsa,~A.~S.; Lebedeva,~I.~V.; Popov,~A.~M.; Knizhnik,~A.~A. Corrigendum to
  ``Long Triple Carbon Chains Formation by Heat Treatment of Graphene
  Nanoribbon: Molecular Dynamics Study with Revised Brenner Potential'' [Carbon
  140 (2018) 543--556]. \emph{Carbon} \textbf{2019}, \emph{146}, 841\relax
\mciteBstWouldAddEndPuncttrue
\mciteSetBstMidEndSepPunct{\mcitedefaultmidpunct}
{\mcitedefaultendpunct}{\mcitedefaultseppunct}\relax
\EndOfBibitem
\bibitem[Sinitsa \latin{et~al.}(2020)Sinitsa, Lebedeva, Polynskaya, Popov, and
  Knizhnik]{Sinitsa2020}
Sinitsa,~A.~S.; Lebedeva,~I.~V.; Polynskaya,~Y.~G.; Popov,~A.~M.;
  Knizhnik,~A.~A. Molecular Dynamics Study of sp-Defect Migration in Odd
  Fullerene: Possible Role in Synthesis of Abundant Isomers of Fullerenes.
  \emph{J. Phys. Chem. C} \textbf{2020}, \emph{124}, 11652--11661\relax
\mciteBstWouldAddEndPuncttrue
\mciteSetBstMidEndSepPunct{\mcitedefaultmidpunct}
{\mcitedefaultendpunct}{\mcitedefaultseppunct}\relax
\EndOfBibitem
\bibitem[Berendsen \latin{et~al.}(1984)Berendsen, Postma, van Gunsteren,
  DiNola, and Haak]{Berendsen1984}
Berendsen,~H. J.~C.; Postma,~J. P.~M.; van Gunsteren,~W.~F.; DiNola,~A.;
  Haak,~J.~R. Molecular Dynamics with Coupling to an External Bath. \emph{J.
  Chem. Phys.} \textbf{1984}, \emph{81}, 3684--3690\relax
\mciteBstWouldAddEndPuncttrue
\mciteSetBstMidEndSepPunct{\mcitedefaultmidpunct}
{\mcitedefaultendpunct}{\mcitedefaultseppunct}\relax
\EndOfBibitem
\bibitem[Smith and Luzzi(2001)Smith, and Luzzi]{Smith2001}
Smith,~B.~W.; Luzzi,~D.~E. Electron Irradiation Effects in Single Wall Carbon
  Nanotubes. \emph{J. Appl. Phys.} \textbf{2001}, \emph{90}, 3509--3515\relax
\mciteBstWouldAddEndPuncttrue
\mciteSetBstMidEndSepPunct{\mcitedefaultmidpunct}
{\mcitedefaultendpunct}{\mcitedefaultseppunct}\relax
\EndOfBibitem
\bibitem[Krasheninnikov \latin{et~al.}(2005)Krasheninnikov, Banhart, Li,
  Foster, and Nieminen]{Krasheninnikov2005}
Krasheninnikov,~A.~V.; Banhart,~F.; Li,~J.~X.; Foster,~A.~S.; Nieminen,~R.~M.
  Stability of Carbon Nanotubes under Electron Irradiation: Role of Tube
  Diameter and Chirality. \emph{Phys. Rev. B} \textbf{2005}, \emph{72},
  125428\relax
\mciteBstWouldAddEndPuncttrue
\mciteSetBstMidEndSepPunct{\mcitedefaultmidpunct}
{\mcitedefaultendpunct}{\mcitedefaultseppunct}\relax
\EndOfBibitem
\bibitem[Banhart \latin{et~al.}(2005)Banhart, Li, and
  Krasheninnikov]{Banhart2005}
Banhart,~F.; Li,~J.~X.; Krasheninnikov,~A.~V. Carbon Nanotubes under Electron
  Irradiation: Stability of the Tubes and Their Action as Pipes for Atom
  Transport. \emph{Phys. Rev. B} \textbf{2005}, \emph{71}, 241408\relax
\mciteBstWouldAddEndPuncttrue
\mciteSetBstMidEndSepPunct{\mcitedefaultmidpunct}
{\mcitedefaultendpunct}{\mcitedefaultseppunct}\relax
\EndOfBibitem
\bibitem[Russo and Golovchenko(2012)Russo, and Golovchenko]{Russoat2012}
Russo,~C.~J.; Golovchenko,~J.~A. Atom-by-Atom Nucleation and Growth of Graphene
  Nanopores. \emph{Proc. Natl. Acad. Sci. U. S. A.} \textbf{2012}, \emph{109},
  5953--5957\relax
\mciteBstWouldAddEndPuncttrue
\mciteSetBstMidEndSepPunct{\mcitedefaultmidpunct}
{\mcitedefaultendpunct}{\mcitedefaultseppunct}\relax
\EndOfBibitem
\bibitem[Warner \latin{et~al.}(2009)Warner, R\"ummeli, Ge, Gemming, Montanari,
  Harrison, B\"uchner, and Briggs]{Warner2009}
Warner,~J.~H.; R\"ummeli,~M.~H.; Ge,~L.; Gemming,~T.; Montanari,~B.;
  Harrison,~N.~M.; B\"uchner,~B.; Briggs,~G. A.~D. Structural Transformations
  in Graphene Studied with High Spatial and Temporal Resolution. \emph{Nat.
  Nanotechnol.} \textbf{2009}, \emph{4}, 500--504\relax
\mciteBstWouldAddEndPuncttrue
\mciteSetBstMidEndSepPunct{\mcitedefaultmidpunct}
{\mcitedefaultendpunct}{\mcitedefaultseppunct}\relax
\EndOfBibitem
\bibitem[Eggen \latin{et~al.}(1996)Eggen, Heggie, Jungnickel, Latham, Jones,
  and Briddon]{Eggen1987}
Eggen,~B.~R.; Heggie,~M.~I.; Jungnickel,~G.; Latham,~C.~D.; Jones,~R.;
  Briddon,~P.~R. Autocatalysis During Fullerene Growth. \emph{Science}
  \textbf{1996}, \emph{272}, 87--90\relax
\mciteBstWouldAddEndPuncttrue
\mciteSetBstMidEndSepPunct{\mcitedefaultmidpunct}
{\mcitedefaultendpunct}{\mcitedefaultseppunct}\relax
\EndOfBibitem
\bibitem[Ewels \latin{et~al.}(2002)Ewels, Heggie, and Briddon]{Ewels2002}
Ewels,~C.~P.; Heggie,~M.~I.; Briddon,~P.~R. Adatoms and Nanoengineering of
  Carbon. \emph{Chem. Phys. Lett.} \textbf{2002}, \emph{351}, 178--182\relax
\mciteBstWouldAddEndPuncttrue
\mciteSetBstMidEndSepPunct{\mcitedefaultmidpunct}
{\mcitedefaultendpunct}{\mcitedefaultseppunct}\relax
\EndOfBibitem
\bibitem[Lebedeva \latin{et~al.}(2008)Lebedeva, Knizhnik,
  Bagatur$^\prime$yants, and Potapkin]{Lebedeva2008}
Lebedeva,~I.~V.; Knizhnik,~A.~A.; Bagatur$^\prime$yants,~A.~A.; Potapkin,~B.~V.
  Kinetics of {2D-3D} Transformations of Carbon Nanostructures. \emph{Physica
  E: Low-dimensional Systems and Nanostructures} \textbf{2008}, \emph{40},
  2589--2595\relax
\mciteBstWouldAddEndPuncttrue
\mciteSetBstMidEndSepPunct{\mcitedefaultmidpunct}
{\mcitedefaultendpunct}{\mcitedefaultseppunct}\relax
\EndOfBibitem
\bibitem[Sun \latin{et~al.}(2006)Sun, Banhart, Krasheninnikov,
  Rodr\'iguez-Manzo, Terrones, and Ajayan]{Sun2006}
Sun,~L.; Banhart,~F.; Krasheninnikov,~A.~V.; Rodr\'iguez-Manzo,~J.~A.;
  Terrones,~M.; Ajayan,~P.~M. Carbon Nanotubes as High-Pressure Cylinders and
  Nanoextruders. \emph{Science} \textbf{2006}, \emph{312}, 1199--1202\relax
\mciteBstWouldAddEndPuncttrue
\mciteSetBstMidEndSepPunct{\mcitedefaultmidpunct}
{\mcitedefaultendpunct}{\mcitedefaultseppunct}\relax
\EndOfBibitem
\bibitem[Ding \latin{et~al.}(2007)Ding, Jiao, Lin, and Yakobson]{Ding2007}
Ding,~F.; Jiao,~K.; Lin,~Y.; Yakobson,~B.~I. How Evaporating Carbon Nanotubes
  Retain Their Perfection? \emph{Nano Lett.} \textbf{2007}, \emph{7},
  681--684\relax
\mciteBstWouldAddEndPuncttrue
\mciteSetBstMidEndSepPunct{\mcitedefaultmidpunct}
{\mcitedefaultendpunct}{\mcitedefaultseppunct}\relax
\EndOfBibitem
\bibitem[Ding \latin{et~al.}(2007)Ding, Jiao, Wu, and Yakobson]{Ding2007a}
Ding,~F.; Jiao,~K.; Wu,~M.; Yakobson,~B.~I. Pseudoclimb and Dislocation
  Dynamics in Superplastic Nanotubes. \emph{Phys. Rev. Lett.} \textbf{2007},
  \emph{98}, 075503\relax
\mciteBstWouldAddEndPuncttrue
\mciteSetBstMidEndSepPunct{\mcitedefaultmidpunct}
{\mcitedefaultendpunct}{\mcitedefaultseppunct}\relax
\EndOfBibitem
\bibitem[Asaka and Kizuka(2005)Asaka, and Kizuka]{Asaka2005}
Asaka,~K.; Kizuka,~T. Atomistic Dynamics of Deformation, Fracture, and Joining
  of Individual Single-Walled Carbon Nanotubes. \emph{Phys. Rev. B}
  \textbf{2005}, \emph{72}, 115431\relax
\mciteBstWouldAddEndPuncttrue
\mciteSetBstMidEndSepPunct{\mcitedefaultmidpunct}
{\mcitedefaultendpunct}{\mcitedefaultseppunct}\relax
\EndOfBibitem
\bibitem[Huang \latin{et~al.}(2005)Huang, Chen, Jo, Wang, Han, Chen,
  Dresselhaus, and Ren]{Huang2005}
Huang,~J.~Y.; Chen,~S.; Jo,~S.~H.; Wang,~Z.; Han,~D.~X.; Chen,~G.;
  Dresselhaus,~M.~S.; Ren,~Z.~F. Atomic-Scale Imaging of Wall-by-Wall Breakdown
  and Concurrent Transport Measurements in Multiwall Carbon Nanotubes.
  \emph{Phys. Rev. Lett.} \textbf{2005}, \emph{94}, 236802\relax
\mciteBstWouldAddEndPuncttrue
\mciteSetBstMidEndSepPunct{\mcitedefaultmidpunct}
{\mcitedefaultendpunct}{\mcitedefaultseppunct}\relax
\EndOfBibitem
\bibitem[Huang \latin{et~al.}(2006)Huang, Chen, Wang, Kempa, Wang, Jo, Chen,
  Dresselhaus, and Ren]{Huang2006}
Huang,~J.~Y.; Chen,~S.; Wang,~Z.~Q.; Kempa,~K.; Wang,~Y.~M.; Jo,~S.~H.;
  Chen,~G.; Dresselhaus,~M.~S.; Ren,~Z.~F. Superplastic Carbon Nanotube.
  \emph{Nature} \textbf{2006}, \emph{439}, 281\relax
\mciteBstWouldAddEndPuncttrue
\mciteSetBstMidEndSepPunct{\mcitedefaultmidpunct}
{\mcitedefaultendpunct}{\mcitedefaultseppunct}\relax
\EndOfBibitem
\bibitem[Lyo and Avouris(1991)Lyo, and Avouris]{Lyo1991}
Lyo,~I.-W.; Avouris,~P. Field-Induced Nanometer- to Atomic-Scale Manipulation
  of Silicon Surfaces with the STM. \emph{Science} \textbf{1991}, \emph{253},
  173--176\relax
\mciteBstWouldAddEndPuncttrue
\mciteSetBstMidEndSepPunct{\mcitedefaultmidpunct}
{\mcitedefaultendpunct}{\mcitedefaultseppunct}\relax
\EndOfBibitem
\bibitem[Chico \latin{et~al.}(1998)Chico, {L\'opez Sancho}, and
  Mu{\~n}oz]{Chico1998}
Chico,~L.; {L\'opez Sancho},~M.~P.; Mu{\~n}oz,~M.~C. Carbon-Nanotube-Based
  Quantum Dot. \emph{Phys. Rev. Lett.} \textbf{1998}, \emph{81},
  1278--1281\relax
\mciteBstWouldAddEndPuncttrue
\mciteSetBstMidEndSepPunct{\mcitedefaultmidpunct}
{\mcitedefaultendpunct}{\mcitedefaultseppunct}\relax
\EndOfBibitem
\bibitem[Liu and Tao(2003)Liu, and Tao]{Liu2003}
Liu,~H.; Tao,~Y. Size Effect of Quantum Conductance in Single-Walled Carbon
  Nanotube Quantum Dots. \emph{Eur. Phys. J. B} \textbf{2003}, \emph{36},
  411--418\relax
\mciteBstWouldAddEndPuncttrue
\mciteSetBstMidEndSepPunct{\mcitedefaultmidpunct}
{\mcitedefaultendpunct}{\mcitedefaultseppunct}\relax
\EndOfBibitem
\bibitem[Ayuela \latin{et~al.}(2008)Ayuela, Jask\'olski, Pelc, Santos, and
  Chico]{Ayuela2008}
Ayuela,~A.; Jask\'olski,~W.; Pelc,~M.; Santos,~H.; Chico,~L. Friedel-Like
  Oscillations in Carbon Nanotube Quantum Dots. \emph{Appl. Phys. Lett.}
  \textbf{2008}, \emph{93}, 133106\relax
\mciteBstWouldAddEndPuncttrue
\mciteSetBstMidEndSepPunct{\mcitedefaultmidpunct}
{\mcitedefaultendpunct}{\mcitedefaultseppunct}\relax
\EndOfBibitem
\bibitem[Triozon \latin{et~al.}(2005)Triozon, Lambin, and Roche]{Triozon2005}
Triozon,~F.; Lambin,~P.; Roche,~S. Electronic Transport Properties of Carbon
  Nanotube Based Metal/Semiconductor/Metal Intramolecular Junctions.
  \emph{Nanotechnology} \textbf{2005}, \emph{16}, 230--233\relax
\mciteBstWouldAddEndPuncttrue
\mciteSetBstMidEndSepPunct{\mcitedefaultmidpunct}
{\mcitedefaultendpunct}{\mcitedefaultseppunct}\relax
\EndOfBibitem
\bibitem[Sloan \latin{et~al.}(2002)Sloan, Kirkland, Hutchison, and
  Green]{Sloan2002}
Sloan,~J.; Kirkland,~A.~I.; Hutchison,~J.~L.; Green,~M. L.~H. Integral Atomic
  Layer Architectures of {1D} Crystals Inserted into Single Walled Carbon
  Nanotubes. \emph{Chem. Comm.} \textbf{2002}, 1319--1332\relax
\mciteBstWouldAddEndPuncttrue
\mciteSetBstMidEndSepPunct{\mcitedefaultmidpunct}
{\mcitedefaultendpunct}{\mcitedefaultseppunct}\relax
\EndOfBibitem
\end{mcitethebibliography}

\end{document}